\documentclass[aps,pre,showpacs]{revtex4}

\usepackage{amssymb,amsmath,amsthm,graphicx}

\newtheorem{theorem}{Theorem}[section]

\theoremstyle{definition}
\newtheorem{definition}[theorem]{Definition}

\newtheorem{corollary}[theorem]{Corollary}
\newtheorem{proposition}[theorem]{Proposition}

\theoremstyle{remark}

\begin{document}

\title{Spreading dynamics on small-world networks with connectivity 
fluctuations and correlations}

\author{Alexei Vazquez\\
The Simons Center for Systems Biology, Institute for Advanced Study\\
Princeton, NJ 08540, USA\\
Department of Physics and Center for Complex Networks Research\\
University of Notre Dame, IN 46556, USA}

\date{\today}

\begin{abstract}

Infectious diseases and computer malwares spread among humans and
computers through the network of contacts among them. These networks are
characterized by wide connectivity fluctuations, connectivity correlations
and the small-world property. In a previous work [A. Vazquez, Phys. Rev.  
Lett. 96, 038702 (2006)] I have shown that the connectivity fluctuations
together with the small-world property lead to a novel spreading law,
characterized by an initial power law growth with an exponent determined
by the average node distance on the network. Here I extend these results
to consider the influence of connectivity correlations which are generally
observed in real networks. I show that assortative and disassortative
connectivity correlations enhance and diminish, respectively, the range of
validity of this spreading law. As a corollary I obtain the region of
connectivity fluctuations and degree correlations characterized by the
absence of an epidemic threshold. These results are relevant for the
spreading of infectious diseases, rumors, and information among humans and
the spreading of computer viruses, email worms and hoaxes among computer
users.

\end{abstract}


\maketitle

\bibliographystyle{apsrev}

\section{Introduction}

Halting an epidemic outbreak in its early stages requires a detailed
understanding of the progression of the number of new infections
(incidence). Current mathematical models predict that the incidence
grows exponentially during the initial phase of an epidemic outbreak
\cite{anderson91,dh00,bbv04,jagers75,mode00}. Within this exponential
growth scenario infectious diseases are characterized by the average
reproductive number, giving the number of secondary infections generated
by a primary case, and the average generation time, giving the average
time elapse between the infection of a primary case and its secondary
cases \cite{anderson91,dh00}. In turn, vaccination strategies are designed
in order to modify the reproductive number and the generation time
\cite{anderson91,dh00,ferguson03,fraser04}.

I have recently shown, however, that this picture dramatically changes
when the graph underlying the spreading dynamics is characterized by a
power law degree distribution \cite{vazquez06a,vazquez06b}, where the
degree of a node is defined as the number of its connections. The
significant abundance of high degree nodes (hubs) carry as a consequence
that most nodes are infected in a time scale of the order of the disease
generation time. Furthermore, the initial incidence growth is no longer
exponential but it follows a power law growth $n(t)\sim t^{D-1}$, where
$D$ is the characteristic distance between nodes on the graph. Yet, these
predictions are limited to uncorrelated graphs and the
susceptible-infected (SI) model.

In this work I extend the theory of age-dependent branching processes
\cite{harris02,jagers75,mode00} to consider the topological properties of
real networks. First, I generalize my previous study
\cite{vazquez06a,vazquez06b} to include degree correlations. This is a
fundamental advance since real networks are characterized by degree
correlations \cite{kr01,pvv01,n02a,ms02} that may significantly affect the
system's behavior \cite{bpv02,vm02,vw02}. Second, I consider the
susceptible-infected-removed (SIR) model that provides a more realistic
description of real epidemic outbreaks \cite{anderson91}, allowing us to
obtain conclusions about the impact of patient isolation and immunization
strategies on the final outbreak size.  Finally, I survey our current
knowledge about different networks underlying the spreading of infectious
diseases and computer malwares and discuss the impact of their topology on
the spreading dynamics.

\section{Population structure}

Consider a population of $N$ susceptible agents (humans, computers, etc)  
and an infectious disease (human disease, computer malware, etc) spreading
among them. The potential disease transmission channels are represented by
an undirected graph, where nodes represent susceptible agents and edges
represent disease transmission channels. For example, when analyzing the
spreading of sexually transmitted diseases the relevant graph is the web
of sexual contacts \cite{liljeros01}, where nodes represent sexually 
active individuals and edges represent sexual relationships.

The degree of a node is the number of edges connecting this node to other
nodes (neighbors) in the graph. Given the finite size of the population
there is a maximum degree $k_{\rm max}$, where $k_{\rm max}$ is at most
$N-1$. I denote by $p_k$ the probability distribution that a node has
degree $k$. The results obtained in this work are valid for arbitrary
degree distributions. Nevertheless, recent studies have shown that several
real networks are characterized by the power law degree distribution

\begin{equation}
p_k = \frac{k^{-\gamma}}{\sum_{s=1}^{k_{\rm max}} s^{-\gamma} }
\label{pkpl}
\end{equation}

\noindent with $\gamma>2$ 
\cite{ba99,fff99,liljeros01,emb02,eckmann04,barrat04}. Therefore, I focus 
the discussion on this particular case.

Real networks are characterized by degree correlations between connected
nodes as well. Networks representing technological and biological systems
exhibit disassortative (negative) correlations with a tendency to have
connections between nodes with dissimilar degrees \cite{pvv01,ms02}. In
contrast, social networks are characterized by assortative (positive)
degree correlations with a tendency to have connections among nodes with
similar degrees \cite{n02a}.  To characterize the degree correlations I
consider the probability distribution $q(k^\prime|k)$ that a neighbor of a
node with degree $k$ has degree $k^\prime$. It is important to note that
the probability distributions $p_k$ and $q(k^\prime|k)$ are related to
each other by the detailed balance condition \cite{boguna03}

\begin{equation}
kp_kq(k^\prime|k) = k^\prime p_{k^\prime} q(k|k^\prime)\ .
\label{db}
\end{equation}

\noindent Although $q(k^\prime|k)$ contains all the information necessary
to characterize the degree correlations it is difficult to analyze. A more
intuitive measure which often appears in the analytical calculations
\cite{bpv02,vm02} is the average neighbors excess degree \cite{pvv01}

\begin{equation}
K_k = \sum_{k^\prime=2}^\infty q(k^\prime|k)(k^\prime -1)\ .
\label{Kk}
\end{equation}

\noindent The empirical data indicates that
\cite{pvv01,vazquez03a,barrat04,batiston04}

\begin{equation}
K_k = c k^{\nu}\ ,
\label{qk}
\end{equation}

\noindent where $c$ is obtained from the detailed balance condition
(\ref{db}), resulting in 

\begin{equation}
c = \frac{ \langle k(k-1)\rangle }{ \langle k^{1+k}\rangle}\ .
\label{c}
\end{equation}

\noindent When the degree correlations are disassortative the nearest
neighbors of a low/high degree node tend to have larger/smaller degree. In
this case $K_k$ decreases with increasing $k$. In contrast, when the
degree correlations are assortative the nearest neighbors of a low/high
degree node tend to have proportional degrees. In this case $K_k$
increases with increasing $k$. Therefore, disassortative and assortative
correlations are characterized by $\nu<0$ and $\nu>0$, respectively.

Real networks also exhibit the small-world property \cite{ws98}, meaning
that the average distance $D$ between two nodes in the graph is small or
it grows at most as $\log N$. For instance, social experiments such as the
Kevin Bacon and Erd\H{o}s numbers \cite{w99} or the Milgram experiment
\cite{m67} reveals that social actors are separated by a small number of
acquaintances. This property is enhanced on graphs with a power law degree
distribution (\ref{pkpl}) with $2<\gamma\leq3$
\cite{chung02,bollobas03,cohen03a}. In this case the average distance
between two nodes grows as $\log\log N$, receiving the name of ultra
small-world \cite{cohen03a}.

Given the graph underlying the spreading of an infectious disease, let us
consider an epidemic outbreak starting from a single node (the root,
patient zero, or index case). In the worst case scenario the disease
propagates to all the nodes that could be reached from the root using the
graph connections. Thus, the outbreak is represented by a spanning or
causal tree from the root to all reachable nodes. The generation of a node
on the tree corresponds with the topological or hopping distance from the
root. This picture motivates the introduction of the following branching
process:

\begin{definition}{Annealed Spanning Tree (AST) with degree correlations}

Consider a graph with degree probability distribution $p_k$ and average
degree $\langle k\rangle$, neighbors degree distribution $q(k^\prime|k)$
given a node with degree $k$, detailed balance condition (\ref{db}), and
average distance between nodes $D$. The annealed spanning tree (AST)
associated with this graph is the branching process satisfying the
following properties:

\begin{enumerate}

\item The process starts from a single node, the root, at generation
$d=0$. The root generates $k$ sons with probability distribution $p_k$.

\item Each son at generation $1\leq d<D$ generates $k^\prime-1$ other sons
with probability distribution $q(k^\prime|k)$, given its parent node has
degree $k$.

\item A son at generation $d=D$ does not generate new sons.

\end{enumerate}

\label{acausaltree}
\end{definition}

\noindent The term annealed means that we are not analyzing the true
(quenched) spanning tree on the graph but a branching process with similar
statistical properties. From the mathematical point of view the AST is a
generalization of the Galton-Watson branching process \cite{harris02} to
the case where (i) the reproductive number of a node depends on the
reproductive number of its ancestor and (ii) the process is truncated at
generation $D$. This mathematical construction has been previously
introduced to analyze the percolation properties of graphs with degree
correlations \cite{n02a}.

The sharp truncation of the branching process at generation $D$ is an
approximation. In the original graph there are some nodes beyond the
average distance between nodes $D$ and their average degree decreases with
increasing generation. Therefore, a more realistic description is obtained
defining $q(k^\prime|k)$ generation dependent \cite{cohen03b,echenique05}
and truncating the branching process when the number of generations equals
the graph diameter. Yet, an analytical treatment of this more realistic
model is either unfeasible or results into equations that most be solved
numerically, questioning its advantage with respect to direct simulations
on the original graph. To allow for an analytical understanding I truncate
the branching process at generation $d=D$, where $D$ represents the
average distance between nodes $D$ in the original graph. Furthermore, I
assume that $q(k^\prime|k)$ is the same for all generations $0\leq d\leq
D$. At this point it is worth noticing that all results derived below are
exact for the AST but an approximation for the original graph.

\section{SIR model of disease spreading}

The AST describes the case where all neighbors of an infected node are
infected and at the same time. More generally a node infects a fraction of
its neighbors and these infections take place at variable times. The
susceptible $\rightarrow$ infected $\rightarrow$ removed (SIR) model is an
appropriate framework to consider the timing of the infection events
\cite{anderson91}. The time scales for the transitions susceptible
$\rightarrow$ infected and infected $\rightarrow$ removed are
characterized by the distribution function of infection and removal times
$G_{\rm I}(\tau)$ and $G_{\rm R}(\tau)$, respectively.  For example,
$G_{\rm I}(\tau)$ is the probability that the infection time is less or
equal than $\tau$.

Consider an infected node $i$ and a susceptible neighbor $j$. The
probability $b(t)$ that $j$ is infected by time $t$ given $i$ was
infected at time zero is the combination of two factors. First, the
infection time should be smaller that $t$ and, second, the removal time of
the ancestor $i$ should be larger than the infection time. More precisely

\begin{equation}
b(t) = \int_0^t dG_{\rm I}(\tau)
\left[1 - G_{\rm R}(\tau)\right]\ .
\label{bt}
\end{equation}

\noindent From $b(t)$ I obtain the probability that $j$ gets infected no
matter when

\begin{equation}
r = \lim_{t\rightarrow\infty}b(t)
\label{r}
\end{equation}

\noindent and the distribution function of the generation times

\begin{equation}
G(\tau) = \frac{1}{r} b(\tau)
\label{Gtau}
\end{equation}

\noindent In the original Kermack-McKendrick formulation of the SIR model
\cite{kermack27} the disease spreads at a rate $\lambda$ from infected to
susceptible nodes and infected nodes are removed at rate $\mu$. In this
case the infection and removal rates $\lambda$ and $\mu$ are exponentially
distributed, $G_{\rm I}(\tau)=1-e^{-\lambda\tau}$ and $G_{\rm R}(\tau) =
1-e^{-\mu\tau}$, resulting in

\begin{equation}
r_{\rm SIR} = \frac{\lambda}{\mu+\lambda}
\label{rSIR}
\end{equation}

\begin{equation}
G_{\rm SIR}(\tau) = 1 - e^{-(\mu+\lambda) \tau}
\label{Gtauexp}
\end{equation}

\noindent Some of the results obtained in this work are valid for any
generation time distribution. We focus, however, on the SIR model with
constant rate of infection and removal (\ref{rSIR})-(\ref{Gtauexp}). 

At this point we can extend the AST definition to account for the variable 
infection times:

\begin{definition}{Age-dependent AST with degree correlations}

The age-dependent AST is an AST where nodes can be in two states,
susceptible or infected, and

\begin{enumerate}

\item An infected node (primary case) infects each of its neighbors 
(secondary cases) with probability $r$.

\item The generation times, the times elapse from the infection of a
primary case to the infection of a secondary case, are independent random
variables with probability distribution $G(\tau)$.

\end{enumerate}

\label{adAST}
\end{definition}

\noindent The age-dependent AST is a generalization of the Bellman-Harris
\cite{harris02} and Crum-Mode-Jagers \cite{jagers75,mode00} age-dependent
branching processes. The key new elements are the degree correlations and
the truncation at a maximum generation, allowing us to consider the
topological properties of real networks.

\section{Spreading dynamics and final outbreak size}

Let $I(t)dt$ be the average number of nodes that are infected between time
$t$ and $t+dt$ given that patient zero (the root) was infected at time
zero. This magnitude is known in the epidemiology literature as the
incidence \cite{anderson91}.  Consider an age-dependent AST and a constant
infection and removal rate (\ref{rSIR})-(\ref{Gtauexp}). Making use of the
tree structure I obtain (Appendix \ref{iterapproach})

\begin{equation}
I(t) = \sum_{d=1}^D z_d \left[
\lambda \frac{(\lambda t)^{d-1}}{(d-1)!} e^{-(\lambda+\mu) t} 
\right]\ ,
\label{ntexp}
\end{equation}

\noindent where

\begin{equation}
z_d = \left\{
\begin{array}{ll}
\langle k\rangle, & d=1\\  
\sum_{k=1}^\infty p_k k(k-1) K_k^{d-2}, & d>1
\end{array}
\right.
\label{zd}
\end{equation}

\noindent is the average number of nodes $z_d$ at
generation $d$, satisfying the normalization condition

\begin{equation}
1 + \sum_{d=1}^Dz_d = N\ .
\label{No}
\end{equation}

\noindent The interpretation of (\ref{ntexp}) is the following. If we
count the time in units of one then on average $z_d$ new nodes are found
at each generation. Since the infection times are variable, however, nodes
at the same generation may be infected at different times.  This
contribution is taken into account by the factor between $[\cdots]$ in
(\ref{ntexp}), giving the probability density of the sum of $d$ generation
times.

Independent of the particular $d$ dependence of $z_d$, from
(\ref{ntexp}) it follows that the incidence decays exponentially for long
times with a decay time $1/(\lambda+\mu)$. This result is a consequence
of the population finite size, i.e. sooner or later most of the nodes are
infected and the number of new infections decays. In contrast, the
factor remaining after excluding the exponential decay is an increasing
function of time and it dominates the spreading dynamics at short and
intermediate times.  I obtain the following result determining the speed
of the initial growth:

\begin{theorem}

Consider the normalized incidence

\begin{equation}
\rho(t) = \frac{I(t)}{N}\ .
\label{rhotnt}
\end{equation}

\noindent If there is some integer $d_c\leq D$ and real numbers $a$ and 
$b>0$ such that for all $d>d_c$

\begin{equation}
\langle p_k k(k-1) K_k^{d-2} \rangle \sim k_{\rm max}^{a+b(d-2)}
\label{cond}
\end{equation}

\noindent when $k_{\rm max}\rightarrow\infty$ then

\begin{equation}
\rho(t) = \lambda 
\frac{(\lambda t)^{D-1}}{(D-1)!} e^{-(\lambda+\mu) t}
\left[ 1 + {\cal O}\left( \frac{t_0}{t} \right) \right]
\ ,
\label{rhot}
\end{equation}

\noindent where

\begin{equation}
t_0 = \frac{1}{\lambda} \frac{D-1}{k_{\rm max}^b}\ .
\label{tau1}
\end{equation}

\label{Nooo}
\end{theorem}

\noindent The symbol ${\cal O}(t_0/t)$ indicates that (\ref{rhot}) is
valid asymptotically when $t\gg t_0$ and it represents correction terms
of the order of $t_0/t$. The demonstration of this result is
straightforward. From (\ref{cond}) it follows that for all $d>d_c$ the
average number of nodes at generation $d$ (\ref{zd}) is of the order of
$z_d\sim k_{\rm max}^{a + b(d-2)}$. Therefore, in the limit $k_{\rm
max}\rightarrow\infty$ the sums in (\ref{ntexp}) and (\ref{No}) are
dominated by the $d=D$ term and corrections are given by the ratio between
the $d=D-1$ and $d=D$ terms.

\begin{figure}

\centerline{\includegraphics[width=4in]{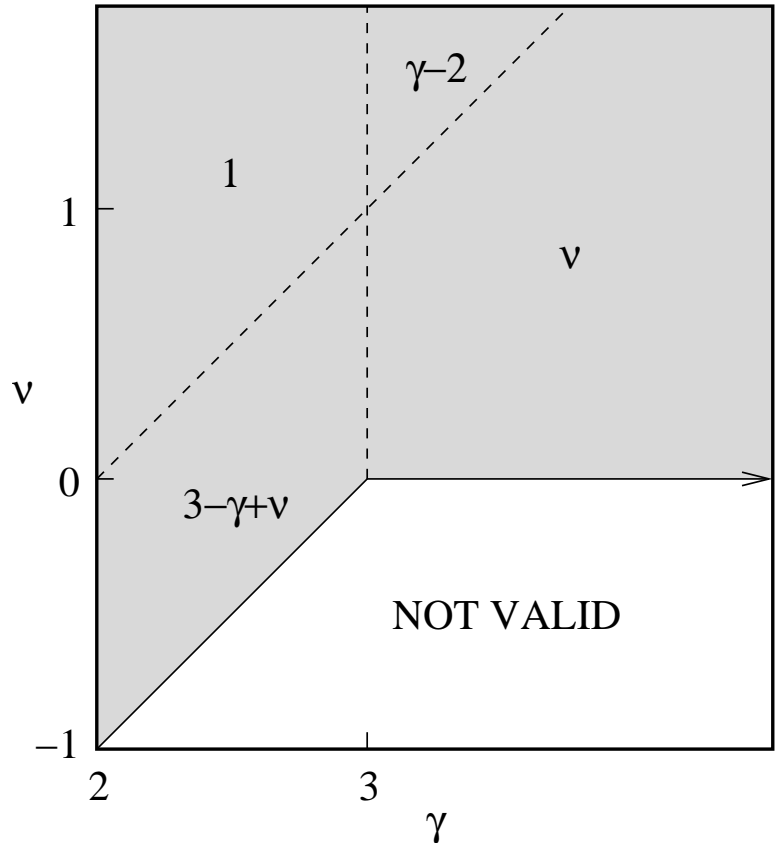}}

\caption{$\gamma-\nu$ plane showing the regions where Theorem \ref{Nooo}
is valid (shadowed region) for the case of a power law degree distribution
(\ref{pkpl}) and degree correlations (\ref{qk}). The text inserts indicate
the exponent $b$ in (\ref{cond}).}

\label{examples}
\end{figure}

The initial dynamics is characterized by a power law growth with an
exponent determined by the average distance $D$. The characteristic time
$t_0$ marks the time scale when this polynomial growth starts to be
manifested. This time is particularly small for graphs with a large
maximum degree and satisfying the small world property, i.e. $D$ is small.  
For instance, let us consider a power law degree distribution (\ref{pkpl})  
with $\gamma>2$ and degree correlations (\ref{qk}). The values of $\gamma$
and $\nu$ for which the condition (\ref{cond}) is satisfied are given in
Fig. \ref{examples}, together with the exponent $b$.  Disassortative
degree correlations ($\nu<0$) may invalidate the condition (\ref{cond}),
indicating that strong disassortative correlations may lead to deviations
from the Theorem \ref{Nooo} prediction. This observation is in agreement
with a previous study focusing on the epidemic threshold \cite{vm02}. In
contrast, for assortative degree correlations ($\nu>0$) the condition
(\ref{cond}) is satisfied for all $\gamma>2$. In other words, assortative
correlations enhance the degree fluctuations, extending the validity of
Theorem \ref{Nooo} to the $\gamma>3$ region.

Focusing on the final size of the outbreak I obtain the following 
corollary:

\begin{corollary}

Consider the average total number of infected nodes

\begin{equation}
N_{\rm I} = N \int_0^\infty dt \rho(t)\ .
\label{NId}
\end{equation}

\noindent If the conditions of Theorem \ref{Nooo} are satisfied then

\begin{equation}
N_{\rm I} = N \left( \frac{\lambda}{\lambda+\mu} \right)^D
\left[ 1 + {\cal O}\left( \frac{(\lambda+\mu)t_0}{D-1} \right)\right]\ .
\label{NI}
\end{equation}

\label{coro}
\end{corollary}

\noindent From this Corollary it follows that increasing the rate of node
removal, because of patient isolation or immunization, we just obtain a
gradual decrease on the final outbreak size. This implies that the concept
of epidemic threshold loses sense since the outbreak size remains
proportional to the population size for all removal rates. This conclusion
is in agreement with previous studies for the case ($\gamma,\nu=0$)
\cite{pv00,n02a,mn00} and ($2<\gamma\leq3,\nu$) \cite{bpv02,vm02}. The
above Corollary extend these studies to the region $\gamma>3$,
demonstrating that when $\nu>0$ there is not an epidemic thresholds for any
value of $\gamma$.

\section{Discussion}

Theorem (\ref{Nooo}) proposes a new law of spreading dynamics
characterized by an initial power law growth. In essence the power low
growth is a consequence of the small-world property and the divergence of
the average reproductive number. Its origin is better understood
analyzing the contribution of nodes at a distance $d$ from the root. The
distribution of infection times of nodes at generation $d$ is given by the
distribution of the sum of $d$ generation times. For the case of a
constant infection rate this distribution is a gamma distribution, which
is characterized by an initial power law with exponent $d-1$. This is the
standard result for stochastic processes defined by a sequence of $d$
steps happening at a constant rate. The total incidence is then obtained
superimposing the contribution of each generation $d$, weighted by the
average number of nodes at that generation. Since most nodes are found at
generation $d=D$ then the contribution from that generation dominates the
incidence progression, resulting in a power law growth with exponent
$D-1$. The small-world properties simply implies that $D$ is small and the
resulting power law growth can be distinguished from an exponential
growth.  The validity of this regime is restricted to time scales that are
large enough such that an appreciable number of nodes at generation $d$
are infected, and it is followed by an exponential decay after most nodes
at that generation are infected.

To understand the relevance of this spreading law for real epidemic
outbreaks, in the following I analyze the validity of the conditions of
Theorem \ref{Nooo} for real networks underlying the spreading of human
infectious diseases and computer malwares.

{\it Sexually transmitted diseases:} Sexual contacts are a dominant
transmission mechanism of the HIV virus causing AIDS. There are several
reports indicating that the web of sexual contacts is characterized by a
power degree distribution.  The current debate is if the exponent $\gamma$
is smaller or larger than three
\cite{liljeros01,jones03a,jones03,schneeberger04}. In either case, it is
known that social networks are characterized by assortative degree
correlations \cite{n02a}, which extends the validity of Theorem \ref{Nooo}
to $\gamma>3$ (see Fig. \ref{examples}). There is also empirical evidence
indicating that the number of AIDS infections grows as a power law in time
for several populations \cite{anderson91,brookmeyer94,szendroi04}. When
this empirical evidence is put together with that for sexual networks we
obtain a strong indication that Theorem \ref{Nooo} provides the right
explanation for the observed power law growth.

{\it Airborne diseases:} Airborne diseases require contact or proximity
between two individuals for their transmission. In this case the graph
edges represent potential contact or proximity interactions among humans
and the degree of an individual is given by the number of people with who
he/she interacts within a certain period of time. Recent simulations of
the Portland urban dynamics \cite{eubank04} shows that the number of
people an individual contacts within a day follows a wide distribution up
to about 10,000 contacts. A report for the 2002-2003 SARS epidemic shows a
wide distribution as well, in this case for the number of secondary cases
generated by a primary SARS infection case. Although this data is not
sufficient to make a definitive conclusion, it provides a clear indication
that the number of proximity contacts a human undergo within a day is
widely distributed. This observation together with the high degree of
assortativity of social networks opens the possibility that the spread of
airborne diseases within a city is described by Theorem \ref{Nooo}.

{\it Computer malwares:} Email worms and other computer malwares such as
computer viruses and hoaxes spread through email communications. The email
network is actually directed, i.e. the observation that user A has user B
on his/her address book does not imply the opposite. This is an important
distinction since the detailed balance condition (\ref{db}) is valid for
graphs with undirected edges. There is, however, a significant
reciprocity, meaning that if user A has user B on his/her address book
then with high probability the opposite takes place as well. Thus, in a
first approximation we can represent email connections by undirected links
or edges and, in such a case, the detailed balance condition (\ref{db})
holds. Recent studies of the email network structure within university
environments indicate that they are characterized by a power law degree
distribution with $\gamma\approx2$ \cite{emb02,eckmann04}. Therefore, the
spreading of computer malwares represents another scenario for the
application of Theorem \ref{Nooo}. Further research is required to
determine the influence of the lack of reciprocity among some email users.

In conclusion, Theorem \ref{Nooo} characterizes the spreading dynamics on
complex networks with wide connectivity fluctuations. Its Corollary
\ref{coro} determines the region of connectivity fluctuations and degree
correlation where there is not an epidemic threshold. The empirical data
indicates that the Theorem conditions are satisfied for several networks
underlying the spreading of infectious diseases among humans and computer
malwares among computers. Therefore, I predict that Theorem \ref{Nooo} is
a spreading law of modern epidemic outbreaks.

\appendix

\section{Iterative approach}
\label{iterapproach}

Let $P_n^{(d,k)}(t)$ be the probability distribution of the number of
infected nodes at time $t$ (including those that has been recovered), $n$,
on a branch of the AST \ref{acausaltree}, given that branch is rooted at a
node at generation $d$ and its has degree $k$. In particular
$P_n^{(0,k)}(t)$ is the probability distribution of the total number of
infected nodes at time $t$, given that patient zero (the root of the tree)
became infected at time zero and its has degree $k$. Based on the tree
structure we can develop an iterative approach to compute $P_n^{(d,k)}(t)$
recursively.

\begin{proposition}

Let $i$ be a node at generation $d$ with degree $k$ and let us denote by
$j$ its neighbors on the next generation $d+1$, where $j\in\{1,\ldots,k\}$
if $d=0$, $j\in\{1,\ldots,k-1\}$ if $0<d<D$, and $j\in\{\emptyset\}$ if
$d=D$. Then

\begin{eqnarray}
\displaystyle
P^{(0,k)}_n(t) & = & 
\sum_{n_1=0}^\infty\dots\sum_{n_k=0}^\infty
\delta_{\sum_{i=1}^kn_i +1, n}
\prod_{i=1}^k 
\sum_{k^\prime=2}^{k_{\rm max}} q(k^\prime|k)
\nonumber\\
& \times &
\left[ r\int_0^t dG(\tau) P^{(d+1,k^\prime)}_{n_i}(t-\tau)
+ \delta_{n_i,0}\left[ 1-r - r[1-G(t)]\right] \right]
\label{PN0}
\end{eqnarray}

\begin{eqnarray}
\displaystyle
P^{(d,k)}_n(t) & = & 
\sum_{n_1=0}^\infty\dots\sum_{n_{k-1}=0}^\infty
\delta_{\sum_{i=1}^{k-1}n_i +1, n}
\prod_{i=1}^{k-1} 
\sum_{k^\prime=1}^{k_{\rm max}} q(k^\prime|k)
\nonumber\\
& \times &
\left[ r\int_0^t dG(\tau) P^{(d+1,k^\prime)}_{n_i}(t-\tau)
+ \delta_{n_i,0}\left( 1-r + r[1-G(t)]\right) \right]
\label{PNd}
\end{eqnarray}

\begin{equation}
P^{(D,k)}_n(t) = \delta_{n,1}\ .
\label{PND}
\end{equation}

\end{proposition}

\begin{proof}

Let $n_i$ be the number of infected nodes in the branch rooted at 
node $i$, and let $n_j$ the number of infected nodes in the 
branches rooted at each of its neighbors $j$. Then

\begin{equation}
n_i = 1 + \sum_j n_j\ .
\label{NiNj}
\end{equation}

\noindent Since node $i$ and its neighbors lie on a tree then $n_j$ are
uncorrelated random variables. Furthermore, all nodes at generation $d$
has the same statistical properties, i.e. $n_j$ are identically
distributed random variables.  Let $Q_n^{(d+1,k)}(t)$ be the probability
distribution of $n_j$, given node $j$ is at generation $d+1$ and its
ancestor has degree $k>0$. With probability $1-r$ the node $j$ is not
infected at any time and with probability $1-G(t)$ it is not yet infected
at time $t$ given it will be infected at some later time.  Thus

\begin{equation}
Q_0^{(d+1,k)}(t) = 1-r + r[1-G(t)]\ .
\label{Q0}
\end{equation}

\noindent On the other hand, with probability $r$ node $j$ will be
infected at some time $\tau$, with distribution function $G(\tau)$, and
continue the spreading dynamics to subsequent generations. Once node $j$
is infected, the number of infected nodes in the branch rooted at node $j$
is a random variable with probability distribution
$P_n^{(d+1,k^\prime)}(t-\tau)$, given node $j$ has degree $k^\prime$.  
Therefore,

\begin{equation}
Q_n^{(d+1,k)}(t) = r\sum_{k^\prime}q(k^\prime|k)\int_0^tdG(\tau)
P_n^{(d+1,k^\prime)}(t-\tau)
\label{QN}
\end{equation}

\noindent for $n>0$. From (\ref{NiNj}), (\ref{Q0}) and (\ref{QN}) we
finally obtain equations (\ref{PN0})-(\ref{PND}).

\end{proof}

Let $I(t)dt$ be the average number of nodes that are
infected between time $t$ and $t+dt$ (incidence), i.e.

\begin{equation}
I(t) = \frac{d}{dt}\sum_{k=0}^\infty p_k 
\sum_{N=0}^\infty P_n^{(0,k)}(t)n\ .
\label{ntd} 
\end{equation}

\noindent Using the recursive relations for $P_n^{(d,k)}(t)$
(\ref{PN0})-(\ref{PND})  we obtain the following result

\begin{proposition}

\begin{equation}
I(t) = \sum_{d=1}^D r^d z_d \frac{dG^{\star d}(t)}{dt}\ ,
\label{nt}
\end{equation}

\noindent where

\begin{equation}
G^{\star d}(t) = \int_0^t dG(\tau_1) \int_0^{\tau_1}dG(\tau_2)\cdots
\int_0^{\tau_{d-1}}dG(\tau_d)
\label{Gd}
\end{equation}

\noindent is the $d$-order convolution of $G(\tau)$, giving the 
probability distribution function of the sum of $d$ generation times.

\label{ntgeneral}
\end{proposition}

\begin{proof}

To obtain $n(t)$ we are going to make use of the generation function 

\begin{equation}
F^{(d,k)}(x,t) = \sum_{n=0}^\infty P_n^{(d,k)}(t) x^n\ .
\label{Fx}
\end{equation}

\noindent Using the recursive equations (\ref{PN0})-(\ref{PND}) for 
$P_n^{(d,k)}(t)$ we obtain

\begin{equation}
F^{(0,k)}(x,t) = x\left[ 1-r + r\left[ 1-G(t) \right] +
r\sum_{k^\prime=1}^{k_{\rm max}} q(k^\prime|k) \int_0^t dG(\tau)
F^{(1,k^\prime)}(x,t-\tau) \right]^k
\label{F0}
\end{equation}

\begin{equation}
F^{(d,k)}(x,t) = x\left[ 1-r + r\left[ 1-G(t) \right] +
r\sum_{k^\prime=1}^{k_{\rm max}} q(k^\prime|k) \int_0^t dG(\tau)
F^{(d+1,k^\prime)}(x,t-\tau) \right]^{k-1}
\label{Fd}
\end{equation}

\begin{equation}
F^{(0,k)}(x,t) = x\ .
\label{FD}
\end{equation}

\noindent We denote by

\begin{equation}
M^{(d,k)}(t) = \frac{\partial F^{(d,k)}(1,t)}{\partial x}
\label{Mdt}
\end{equation}

\noindent the mean number of infected nodes on the branch rooted at a node
at layer $d$ with degree $k$. Making use of the recursive relations
(\ref{F0})-(\ref{FD}) we obtain

\begin{equation}
M^{(0,k)}(t) = 1 + (1-r)\int_0^tdG(\tau)M^{(1,k)}(t-\tau)
\label{M0}
\end{equation}

\begin{equation}
M^{(d,k)}(t) = 1 + (1-r)\int_0^tdG(\tau)M^{(d+1,k)}(t-\tau)
\label{Md}
\end{equation}

\begin{equation}
M^{(D,k)}(t) = 1\ .
\label{MD}
\end{equation}

\noindent Iterating the recursive relations (\ref{M0})-(\ref{MD}) from 
$d=D$ to $d=0$ we obtain

\begin{eqnarray}
M^{(0,k)}(t) & = & 1 + \sum_{d=1}^D r^d G^{\star d}(t)\ ,
k
\sum_{k_1} q(k_1|k)(k_1-1)
\sum_{k_2} q(k_2|k_1)(k_2-1)
\nonumber\\
& \cdots &
\sum_{k_{d-1}} q(k_{d-1}|k_{k-2})(k_{d-1}-1)\ .
\label{M0kt}
\end{eqnarray}

\noindent Note that from (\ref{Fx}) and (\ref{Mdt}) it follows that

\begin{equation}
\sum_{n=1}^\infty P_n^{(0,k)}(t) n = M^{(0,k)}(t)\ .
\label{PNNMkt}
\end{equation}

\noindent Substituting (\ref{M0kt}) into (\ref{PNNMkt}) and the result 
into (\ref{ntd}) we obtain (\ref{nt}) with

\begin{eqnarray}
z_d & = & \sum_k p_k k
\sum_{k_1} q(k_1|k)(k_1-1)
\sum_{k_2} q(k_2|k_1)(k_2-1)
\nonumber\\
& \cdots &
\sum_{k_{d-1}} q(k_{d-1}|k_{k-2})(k_{d-1}-1)\ .
\label{zd1}
\end{eqnarray}

\noindent Finally, using the detailed balance condition (\ref{db}) we 
reduce (\ref{zd1}) to (\ref{zd}).

\end{proof}

This work was supported by NSF Grants No. ITR 0426737 and No. ACT/SGER
0441089.


\end{document}